\documentclass[5p,times]{elsarticle}
\usepackage{amssymb}

\makeatletter
\newcommand{\verbatimfont}[1]{\def\verbatim@font{#1}}%
\makeatother

\usepackage{url}
\usepackage{bm}

\usepackage[table]{xcolor} 

\journal{Parallel Computing}

\begin{document}

\begin{frontmatter}



\title{A comparative evaluation of three volume rendering libraries for the visualization of sheared thermal convection}


\author[CSCS]{Jean M. Favre\corref{cor1}}
\cortext[cor1]{jfavre@cscs.ch}

\author[Twente]{Alexander Blass}

\address[CSCS]{Swiss National Supercomputing Center (CSCS), Via Trevano 131, CH-6900 Lugano, Switzerland}
\address[Twente]{Physics of Fluids Group, Max Planck Center for Complex Fluid Dynamics,
J. M. Burgers Center for Fluid Dynamics and MESA+ Research Institute,
Department of Science and Technology,
University of Twente, P.O. Box 217, 7500 AE Enschede, The Netherlands}

\begin{abstract}
Oceans play a big role in the nature of our planet, about $ 70 \% $ of our earth
is covered by water \cite{int14}. Strong currents are transporting warm water around the world making life possible, and allowing us to harvest its
power producing energy. Yet, oceans also
carry a much more deadly side. Floods and tsunamis can easily annihilate whole
cities and destroy life in seconds. The earth's climate system is also very much
linked to the currents in the ocean due to its large coverage of the earth's surface, thus, gaining scientific insights into the mechanisms and effects through simulations is of high importance.
Deep ocean currents can be simulated by means of wall-bounded turbulent flow simulations.
To support these very large scale numerical simulations and enable the scientists to interpret their output,
we deploy an interactive visualization framework to study sheared thermal convection.
The visualizations are based on volume rendering of the temperature field.
To address the needs of supercomputer users with different hardware and software resources,
we evaluate different volume rendering implementations supported in the ParaView \cite{ahr05} environment:
two GPU-based solutions with Kitware's native volume mapper or NVIDIA's IndeX library,
and a CPU-only Intel OSPRay-based implementation.

\end{abstract}


\end{frontmatter}


\begin{figure}[!hbt]
	\centering
	\includegraphics[width=\linewidth]{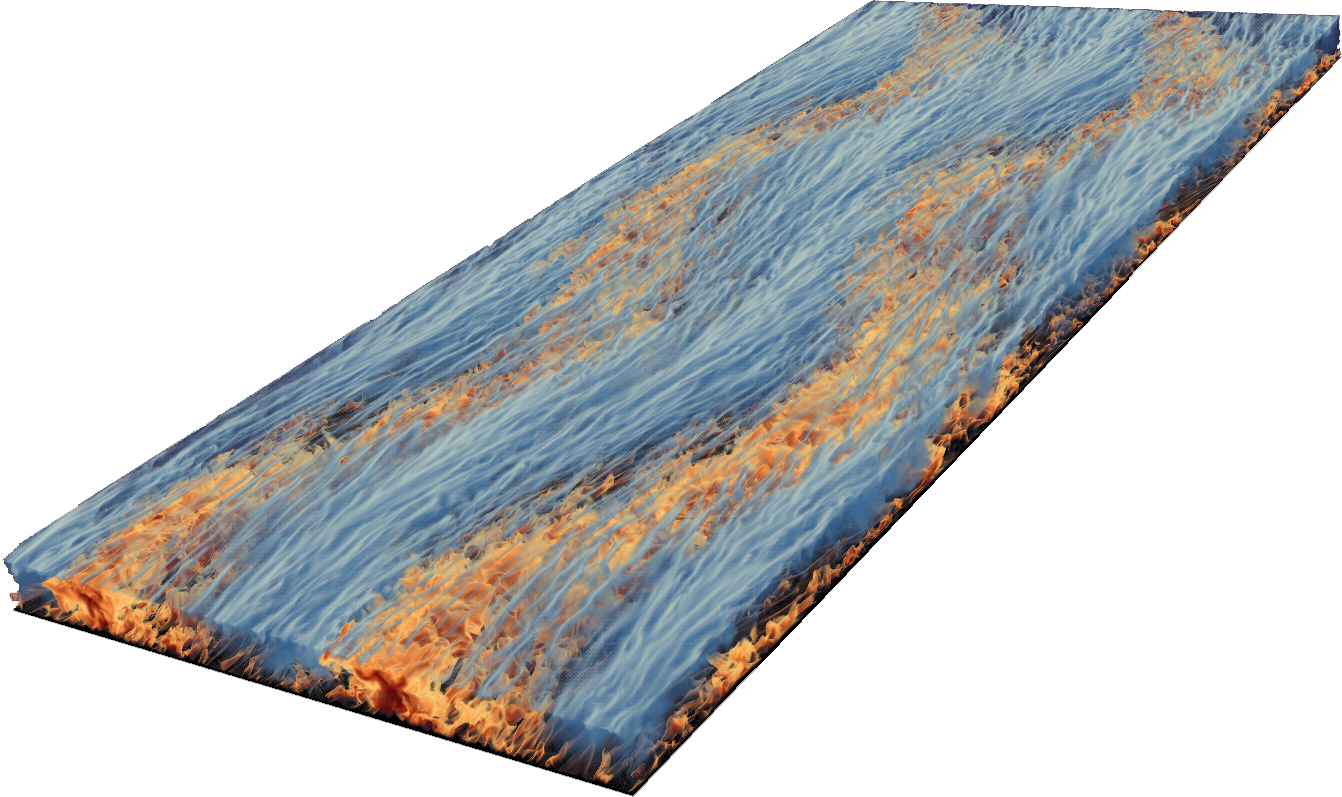}%
	\caption{\label{fig:flowfield} Snapshot of the three-dimensional temperature field of sheared thermal convection at $ Ra=4.6 \times 10^6 $ and $ Re_w=6000 $ \cite{bla19}.}
\end{figure}

\section{Introduction}
\label{sec:Introduction}

Thermohaline ocean circulation \cite{rah00} is vital for the heat budget of our earth. Manabe and Stouffer \cite{man88} observed that it can contribute to a heat increase of up to $ \sim 10 ^ \circ $C on the yearly averaged mean surface temperatures in the North Atlantic region. Marshall and Schott \cite{mar99} investigated a vast variety of scales in ocean dynamics and stated that deep convection can be related to mixing layers everywhere in the ocean. Since there are many complex three-dimensional events happening in large-scale fluid bodies such as oceans, it is vital to visualize the three-dimensional and temporal features of such flow simulations.

We study these large-scale bodies of fluids which are sheared by winds or currents and influenced by temperature differences in the flow. A fundamental setup of this natural mechanism is sheared thermal convection (Fig. \ref{fig:flowfield}). Many processes in nature are based on heat and momentum transfer and therefore interaction between buoyancy \cite{ahl09,loh10} and shear \cite{smi11,bar07}. Rayleigh-B\'enard convection, the flow in a box heated from below and cooled from above is a paradigmatic system for thermal convection. We present the use of three different rendering libraries available in ParaView \cite{ahr05} to build a time-dependent volume rendering of thermal convection. The deployment and evaluation of the hardware and software requirements of these libraries was motivated by a showcase submission at the 2018 International Conference for High Performance Computing, Networking, Storage and Analysis. In the accompanying video \cite{fav18a} we are able to display the previously two-dimensionally presented flow structures in a three-dimensional motion. The reader is led through a presentation of one specific flow case with sheared thermal convection and can experience the dynamics of the thermal structures while being informed about different flow parameters.

\section{Numerical simulations of sheared thermal convection}
The direct numerical simulations (DNS) were performed with the second-order finite-difference
code \textit{AFiD} \cite{poe15c}, in which the three-dimensional non-dimensional
Navier-Stokes equations with the Boussinesq approximation are solved on a staggered grid.

%
%

We use $ \boldsymbol{u}=u(\boldsymbol{x},t) $ as the velocity vector with streamwise, spanwise and wallnormal components. $ \theta $ is the non-dimensional temperature ranging from $ 0 \leq \theta \leq 1 $. The simulations are performed in a computational box with periodic boundary conditions in streamwise and spanwise directions and confined by a heated plate below and a cooled plate on top. The shearing of the flow is implemented by a Couette flow setting where both top and bottom plates of the flow are moved in opposite directions with the speed $ u_w $ keeping the average bulk velocity at zero and therefore minimizing dissipation errors. The domain size is ($ L_x \times L_y \times L_z $) = ($ 9\pi h \times 4\pi h \times h $) using a grid of ($ n_x \times n_y \times n_z $) = ($ 6912 \times 3456 \times 384 $) which is homogeneously distributed in the streamwise and spanwise directions and clustered towards the walls.

The open source finite-difference Navier-Stokes solver \textit{AFiD} \cite{poe15c} was written in Fortran 90 to study large-scale wall bounded turbulent flow simulations. In collaboration with NVIDIA, USA, the code was ported in its newest version to a GPU setting using an MPI and CUDA Fortran hybrid implementation optimized to run and solve large flow fields \cite{zhu18b}.

We used data from Blass et al. \cite{bla19} for our evaluation of volume rendering implementations, where a parameter study over different input parameters was conducted to study their influence on the flow field. Control parameters were the temperature difference between the top and bottom plates as the strength of the thermal forcing, non-dimensionalized as the Rayleigh number $Ra$, and the wall velocity as the strength of the shear forcing, non-dimensionalized as the wall shear Reynolds number $Re_w$.


%
%


\begin{figure}
	\centering
	\includegraphics[width=0.8\linewidth]{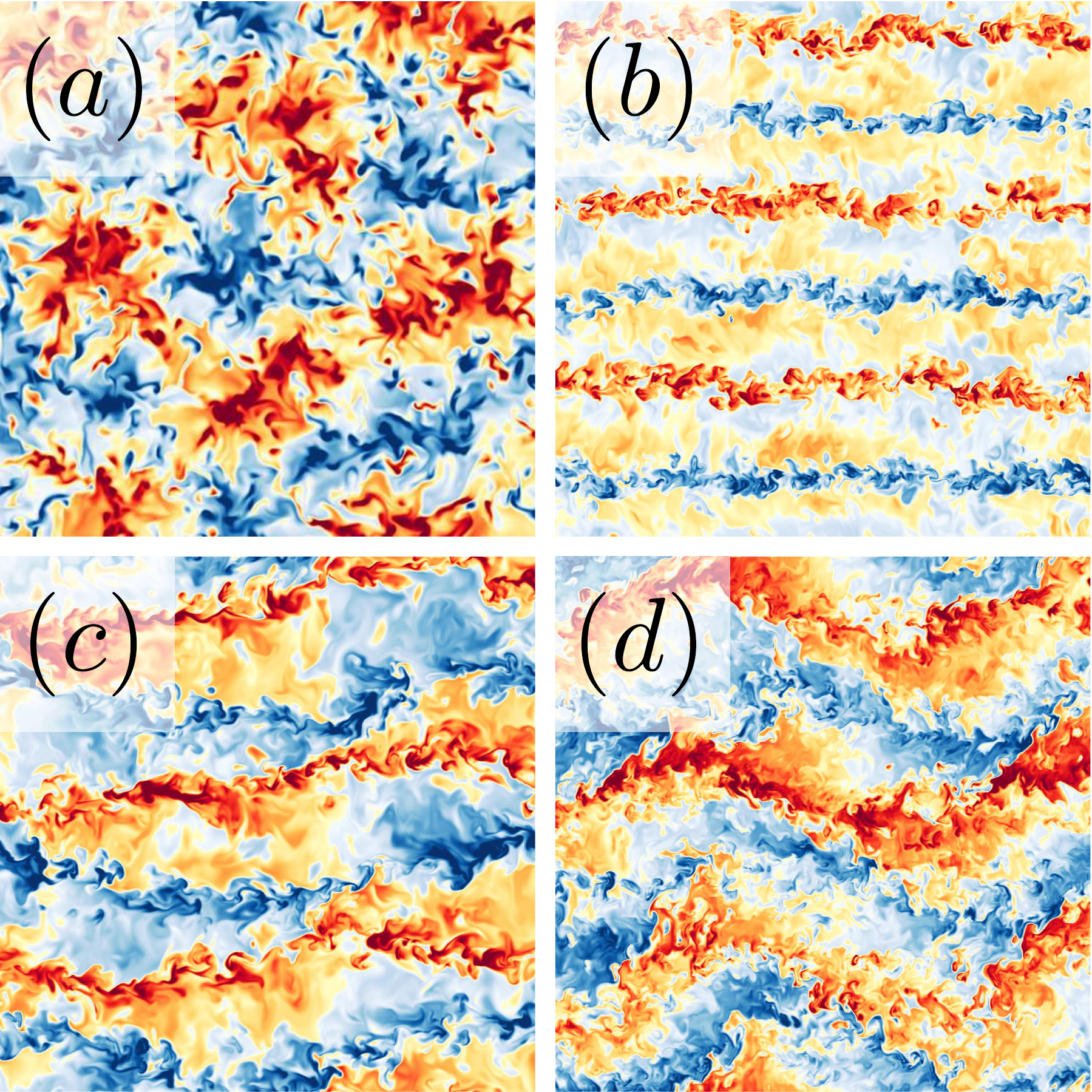}%
	\caption{\label{fig:overview} Zoomed snapshots of temperature fields
of a sheared and thermally forced flow transitioning through all flow regimes
for $ Ra=2.2 \times 10^6 $ and (a) $ Re_w=0 $, (b) $ Re_w=2000 $, (c) $ Re_w=3000 $, (d) $ Re_w=6000 $, ranging from $ \theta_{min} $ (blue) to $ \theta_{max} $ (red).}
\end{figure}

In Fig. \ref{fig:overview} we present snapshots of temperature
fields at mid height in different flow regimes. It can be observed that the flow passes from a
thermally dominated regime with large thermal convection rolls driving the flow (Fig. \ref{fig:overview}a) into a
regime where the mechanical forcing is dominant. Here, large-scale meandering
structures can be observed which are driven by the shearing of the top and bottom plates (Fig. \ref{fig:overview}d). To undergo a transition between the
regimes, the flow has to pass through an intermediate stage, in which the thermal plumes
get stretched into large streaks (Fig. \ref{fig:overview}b). If the shearing is
further increased, these streaks become unstable and start meandering in the
final flow state (Fig. \ref{fig:overview}c,d).

The reason for this streaky flow behavior is the addition of a third dimension to originally quasi-two-dimensional flow structures in pure thermal convection. Such thermal convection rolls are driven solely by the thermal difference between the plates. Once the wall shearing is added, the flow starts to strongly move in streamwise direction, which causes the development of streaks.

In turbulent flows it is very important to research how certain characteristic
parameters are influenced by the flow. In thermal convection, the heat transfer, non-dimensionally
defined through the Nusselt number $Nu$ is a good indicator if changing flow structures have a supporting effect or may disrupt a previously transport-favorable flow situation. 


While two-dimensional visualizations are very helpful in understanding the behavior of the large-scale structures, they don't show the complete scientific picture. They give a good indication of the flow behavior, but to understand thermal turbulence, it is vital to see the whole flow field and the dynamic interaction of turbulent structures with each other.
The opportunity to observe the flow evolving and transitioning through different regimes is a great chance to not only statically observe different flow states at fixed locations in space, but to also actually follow the flow on its path to develop thermal plumes, streaks and meandering structures.

\begin{figure}
	\centering
	\includegraphics[width=\linewidth]{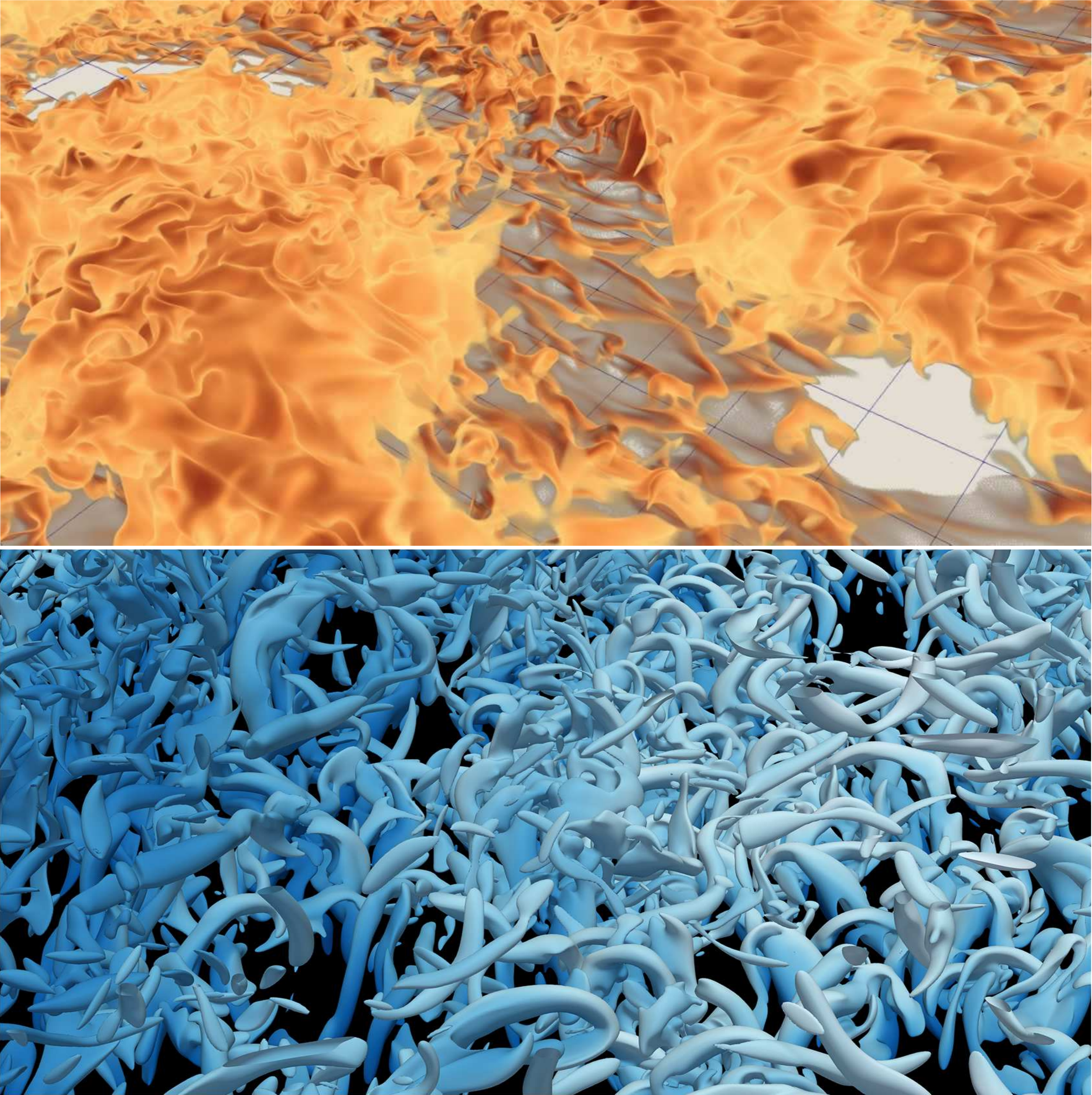}%
	\caption{\label{fig:smallscale} Zoom of an snapshot of the temperature field (top) and the vorticity structures (bottom) at $ Ra=2.2 \times 10^6 $ and $ Re_w=6000 $.}
\end{figure}

It has been previously shown in thermal convection that the large thermal plumes can be traced until very close to the heated and cooled plates \cite{ste18}. So it is very important to also observe the emergence of structures close to the boundary layer. In the shear dominated regime, which we visualize in the accompanying video \cite{fav18a}, we can observe extremely large-scale structures which are caused by a combination of thermal and shear forcing. The detailed visualizations we presented allow us to not only follow the large-scale structures, but also the interaction of small-scale structures much closer to the plates (Fig. \ref{fig:smallscale}).

\section{Volume rendering libraries and setup}

We use ParaView v5.6.0, a world-class, open source, multi-platform data analysis and
visualization application installed on Piz Daint. Piz Daint, a hybrid Cray XC40/XC50 system,
is the flagship supercomputer of the Swiss National HPC service. We have deployed
and tested several solutions within ParaView where parallelism is expressed
at different degrees: data-parallel visualization pipelines with GPU-based renderings
or multi-threaded parallelism for CPU-based renderings.

The computational domain used for our simulations is made of $ 6912 \times 3456 \times 384 $ grid points.
The temperature scalar field stored as \it{float32} \rm takes 36 GB of memory, an
overwhelming size to handle on a normal desktop. Using different parallel programming
paradigms has enabled us to provide an engaging environment to promote interactive tuning of
visualization options and high productivity for movie generation.

Visualization of three-dimensional scalar fields is a very mature field. Many techniques are
available to make some sense of the three-dimensional nature of the data, and its variations
throughout the volume. Surface-based renderings with isosurface thresholds or
slicing planes have a great appeal in that they are easy to use, and provide unambiguous
representations based on clearly defined numerical values. Volume renderings, early applied
to medical applications, are also a great fit for scalar visualizations, especially in the
realm of time-dependent outputs. They are, however, much more difficult to use. Volume rendering is
based on the principle of converting a 3D scalar field onto an RGB (color) volume and an Opacity volume.
Transfer functions, often defined in an ad-hoc manner, convert scalar values to colors, and classify
the data into regions of different opacities. A volume can then appear as clouds with varying density and color.
Their interpretation remains subjective to the user's taste and practice.
We refer readers to other sources \cite{sch06} to dive more deeply into the principles of Volume Rendering.

Volume Rendering can be implemented in different manners. ParaView was chosen because it offers a testbed
for several state of the art implementations which can be selected based on rendering parameters and
available hardware.

The largest partition of the Piz Daint supercomputer has nodes equipped
with one Intel Xeon E5-2690 (12 cores, 64 GB RAM) and one NVIDIA
Tesla P100 GPU (16 GB RAM, OpenGL driver 396.44). Thus our priority is to evaluate
the GPU-based implementations.
ParaView's default installation enables also a software ray caster for rendering volumes but we have found its performance far
below the other options. The lack of advanced parameter settings in the Graphical
User Interface (GUI) of ParaView also led us to abandon its evaluation. We tested ParaView's native GPU
ray casting implementation against IndeX an NVIDIA library, as well as OSPRay, a software-based library developed by Intel.
Doing so, provides a valid option to users of supercomputers not equipped with GPUs.
Our performance evaluation is based on ParaView's benchmarking Python source
code\footnote{source code found in ./Wrapping/Python/paraview/benchmark/}.

We have in all cases ignored disk-based I/O costs. There is often quite
a bit of variability when running on a large distributed file system shared by
hundreds of users. Our motivations are rendering-centered, and two-fold:
evaluate the memory cost and resources (CPU, GPU) required to get a first image
on the screen, and see if color/opacity transfer
function editing, as well as other image tuning, can be done interactively, using
any of the three methods proposed. In the evaluation of performance costs, ParaView's
benchmark code enables fully automated testing with a careful management of
double buffering, turning off all rendering optimizations designed to accelerate
interactive viewing, and forcing full-feature rendering before saving images to disk.

In the two GPU-based methods evaluated, we use an EGL-based rendering layer \cite{egl19} to
overcome the need to have a server-side X-Windows server running on the compute node.
This enables headless, offscreen rendering with GPU acceleration. We note, however, that although
the GPUs provide phenomenal rendering power, they are limited by the available memory
(16 GB on our NVIDIA's Pascal GPUs). For the full size of our simulations outputs,
we are actually forced to use data-parallel pipelines on multiple nodes to use
the aggregate memory of the different GPUs.

Our third option, uses Intel OSPRay and CPU rendering. HPC compute nodes usually
have more memory than their GPU counterparts. We use Piz Daint's high memory nodes
with 128 GB of RAM, where our grid of over 9 billion voxels can be fit
easily on a single node.

\subsection{ParaView's GPU ray casting} \label{smart}

When GPU hardware is present, ParaView's most efficient mapper is a volume
mapper that performs ray casting on the GPU using vertex and fragment programs \cite{kit19}.
The core ray-tracing algorithms are coded in GLSL and require a graphics driver
supporting at least OpenGL version 3.2 \cite{sha19}. The data is stored
into a vtkVolumeTexture which manages the OpenGL volume texture, its type and
internal format. Although this class supports streaming data into separate blocks
to make it fit the GPU memory, we have not used this option which imposes
a performance trade-off, artificially going over the fixed GPU memory limit.
Block streaming, sometimes called data bricking, may also suffer from artifacts at
the block boundaries where gradient computations are done to support shading.
ParaView's OpenGL VolumeRayCastMapper binds the 32-bit float scalar field array
to a three-dimensional texture image with a call to glTexImage3D().
An explicit texture object is created, transferring data from host memory to GPU memory.
The maximum achievable performance will be proportional to the total amount of GPU memory,
and to the transfer bandwidth over our high speed PCIe3 serial bus connecting the host to the GPU device.

\begin{figure}
	\centering
	\includegraphics[width=\linewidth]{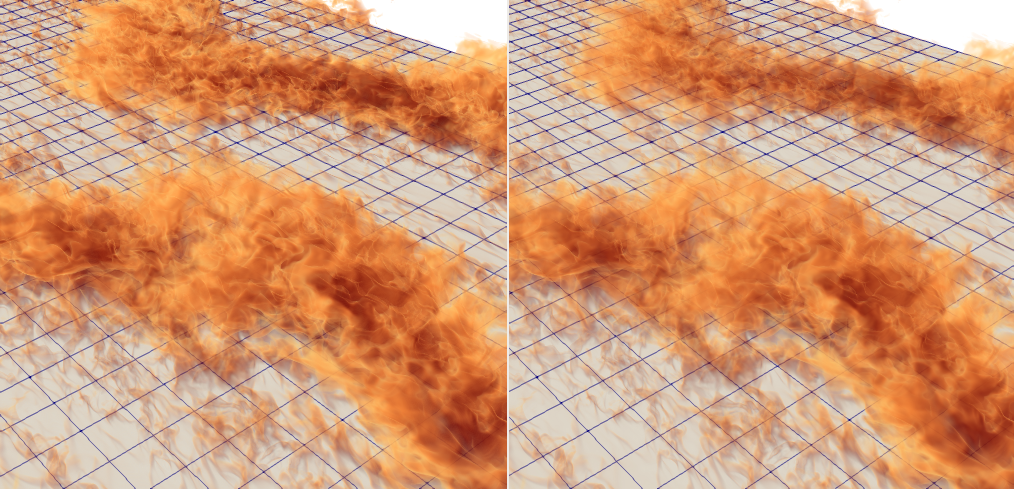}%
	\caption{\label{fig:gpucloseup} Comparison between volume renderings of temperature with ParaView's
OpenGL GPU RayCastMapper (left), and with NVIDIA IndeX (right).}
\end{figure}

\subsection{NVIDIA IndeX} \label{index}

NVIDIA IndeX \cite{nvi19} is a three-dimensional visualization SDK developed to enable
volume rendering of massive data sets. NVIDIA has worked in tandem with Kitware to
bring an implementation of IndeX to ParaView, and we have enjoyed the benefits
of a close partnership between the Swiss National Supercomputing Center (CSCS)
and NVIDIA, to be able to use IndeX in a multi-GPU setting. We use the ParaView
plugin v2.2 with the core library NVIDIA IndeX 2.0.1.
The NVIDIA IndeX Accelerated Compute (XAC) interface integrates the core surface
and volume sampling programs written in CUDA \cite{haa18}.
For this case, we have used the generic programs provided by IndeX, without custom programming.
In Fig.~\ref{fig:gpucloseup} we show side-by-side renderings done with the
two GPU-based libraries, to demonstrate that they produce equivalent images.
The ParaView Graphical User Interface ensures that both implementations use identical
color and opacity transfer functions and sampling rates. ParaView's
GPU Ray Casting image (left) is used as reference. Differences of
illumination are barely noticeable to the human eye.

\subsection{Intel OSPRay}

OSPRay \cite{osp19} is a ray tracing framework for CPU-based rendering. It supports advanced 
shading effects, large models and non-polygonal primitives. OSPRay can distribute 
``bricks'' of data as well as ``tiles'' of the framebuffer, although in our case, we use brick subdivisions only. The Texas Advanced Computing Center
has developed a ParaView plugin that enables us to test the possibility of
using a ray-tracing based rendering engine for volumetric rendering. This is
the best solution for clusters where no GPU hardware is available.

OSPRay can use its own internal Message Passing Interface (MPI) layer to replicate
data across MPI processes and composite the image. This would result in linear
performance scaling and supports secondary rays used in ParaView's \it{pathtracer} \rm mode,
but would be prohibitive in terms of communication costs.
In this study, we rely on a different parallel computing paradigm.
The emphasis is no more on data parallelism, but rather on multi-threaded execution.
A complete \it{software-only} \rm ParaView installation was deployed with an LLVM-based OpenGL Mesa layer. We used Mesa v17.2.8, compiled with LLVM v5.0.0, and the
OSPRay v1.7.2 library to provide a very efficient multi-threaded execution path
taking advantage of Piz Daint's second partition of compute nodes. These nodes
are built with two Intel Broadwell CPUs (2x18 cores and 64/128 GB RAM). Our
cluster management and job scheduling system SLURM provides the specific scheduling options
``\texttt{--}cpus-per-task=72 \texttt{--}ntasks-per-core=2'' to effectively
take full advantage of the multi-threading exposed by the LLVM and OSPRay libraries. 

\subsection{Parallel image compositing}

ParaView's default mode of parallel computing is to use data-parallel distribution,
whereby sub-pieces of a data grid are processed through identical visualization
pipelines. To combine the individual framebuffers of each computing nodes,
ParaView uses Sandia National Laboratory's IceT \cite{mor11} compositing
library. We use it in its default mode of operation doing sort-last compositing
for desktop image delivery. We note here that NVIDIA's IndeX uses a proprietary
compositing library, so for the IndeX tests only, we disable ParaView's default
image compositor.

\section{Volume rendering of the thermal convection}

\begin{figure}[!hbt]
	\centering
	\includegraphics[width=\linewidth]{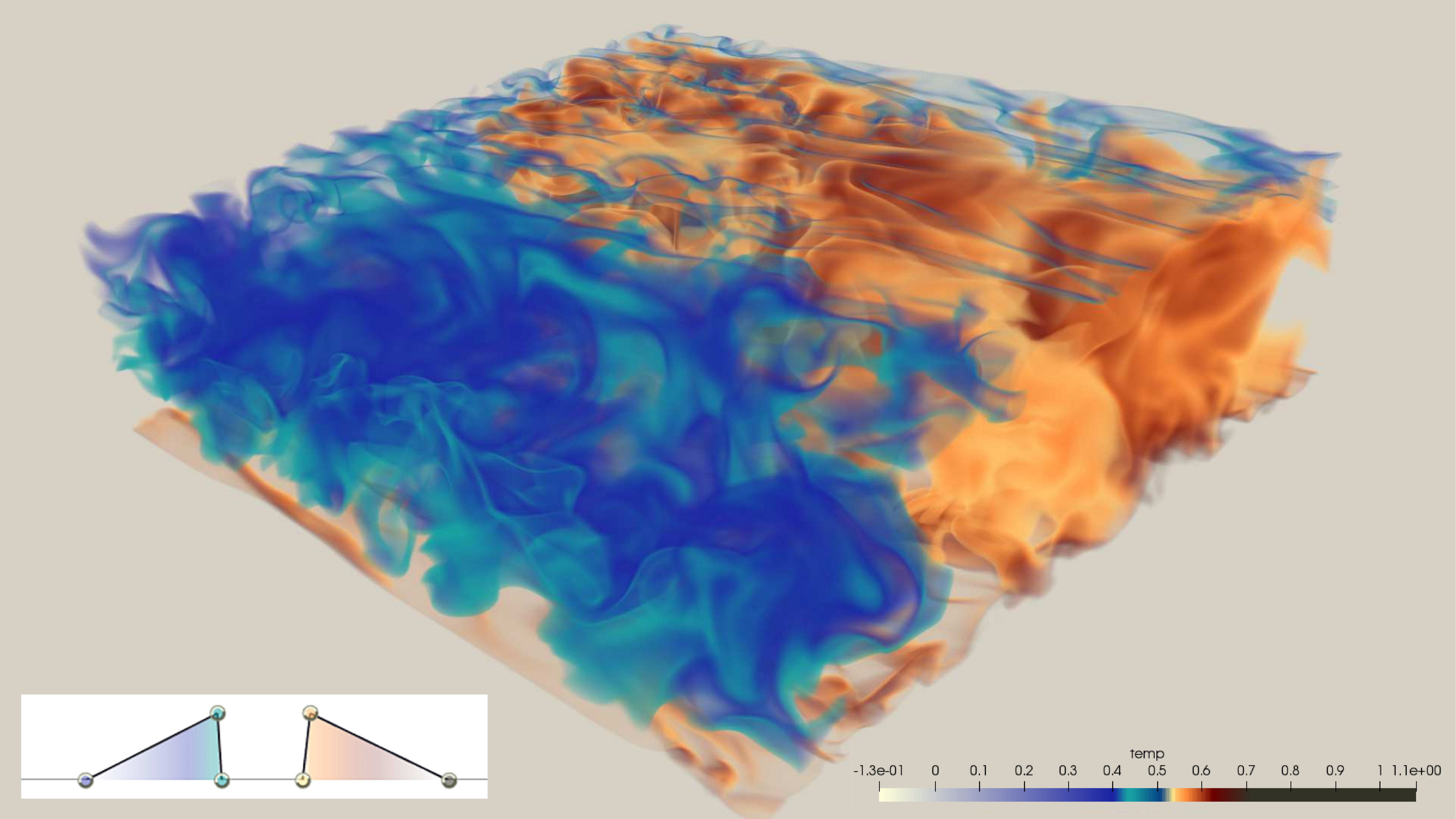}%
	\caption{\label{fig:zoom} Example of a color and opacity transfer functions to highlight hot and cold plumes.}
\end{figure}

In visualizing the temperature field, we seek to highlight the turbulence which
is best shown by clearly differentiating between cold and hot regions to see how
they interact with each other, as seen in Fig.~\ref{fig:zoom}. Our movie animation
shows an initial phase where region of blue tint is superposed on top of the hotter
region. Plumes emerging from the bottom and mixing into the cold regions highlight
this phenomenon.

\subsection{Visual effects}

\begin{figure}
	\centering
	\includegraphics[width=\linewidth]{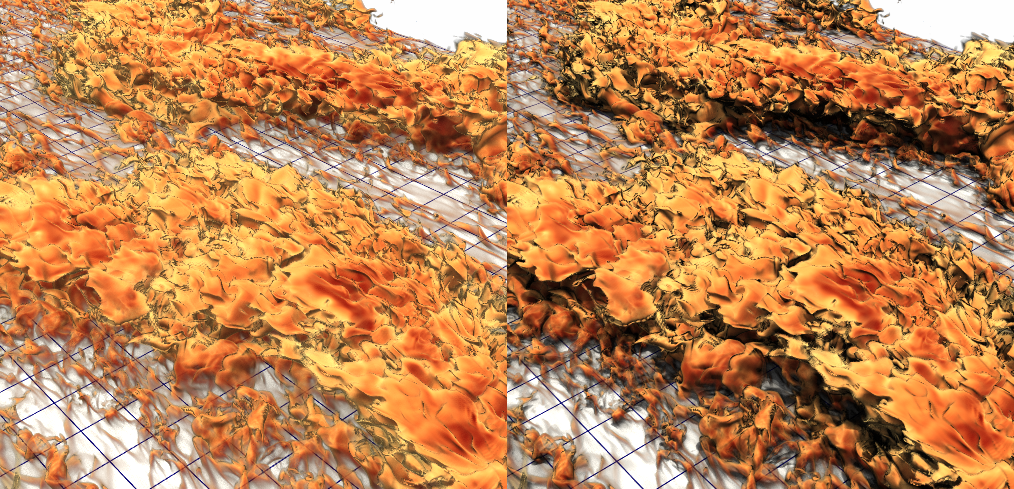}%
	\caption{\label{fig:shadings} Volume rendering with shading based on gradient
estimation (left), and with OSPRay-enabled shadows (right).}
\end{figure}

When presented with multiple visualizations including different illumination and
shading, we preferred the renderings which emphasize the
amorphous nature of the field data. As can be seen in Fig.~\ref{fig:shadings},
shading based on gradient estimation offers little improvement because our data
does not have strong gradients, and the use of shadows which at first might seem
more appealing, produces images with a strong \it{surface-like} \rm look, which
we discarded upon further analysis.

\subsection{GPU-based rendering on a single node}

Volumetric rendering of high resolution grids has a non-significant cost which we
briefly document here. Creating the first frame after data has been read in memory,
 i.e., the startup cost has a great impact in having users adopt a particular implementation.
In a \it{post-hoc} \rm visualization, data would be read from disk; in an \it{in-situ} \rm
scenario, data might have to be converted to VTK data structures. Thus, we measure performance
after the time ParaView has collected all the data and created a bounding-box representation.
This startup cost for the first image is also of paramount importance in a movie-making scenario,
where data are read from disk, a single image is computed, and the whole visualization pipeline
and hardware resources are flushed to visualize the next timestep.

Unlike ParaView's native GPU ray caster implementation which does not enable block
streaming, the NVIDIA IndeX library processes data by chunks. However, it does so
by bringing volume sub-extents \it{incrementally} \rm into the GPU memory.
Early volume chunks are rendered properly as long as the GPU memory is not exhausted.
When memory runs out, late chunks actually corrupt the final image. Our attempts to
render a 4 billion voxels dataset on a single node did not succeed with NVIDIA IndeX.
We observe failures to allocate $64^3$ voxel cubes and the final images are corrupted.

We summarize in Table \ref{tab:firstframe-tab} the time from when volumetric rendering options
are enabled, triggering the building of internal structures until the first frame appears.
In order to measure the memory cost of all three libraries under evaluation on a single node,
we restricted our test sample to a quarter-size domain of the original grid, i.e., 2.28G voxels
($ 1730 \times 3456 \times 384 $), to fit the available GPU RAM. The GPU memory
usage\footnote{GPU memory usage is measured with the nvidia-smi diagnostic tool}
settles at 9.1 GB for ParaView native raycaster, and 12.3 GB for NVIDIA IndeX. 

\begin{table}[htb]
  \centering
  \caption{
    Initialization and memory costs for a quarter-size domain on one node.
  }
  \label{tab:firstframe-tab}

  \begin{tabular}{lccc}
    \hline
    Rendering library & Startup & ParaView task\\
    \hline
    OSPRay & 1.34 s &  18.4 GB \\
    ParaView GPU Mapper & 6.17 s &  27.2 GB \\
    NVIDIA IndeX & 11.84 s &  39.2 GB\\
    \hline

  \end{tabular}
\end{table}

We note both a much higher memory consumption on the application side of ParaView
and on the GPU memory side for the NVIDIA IndeX implementation. The high initial 
setup cost incurred by the NVIDIA IndeX library is due to higher volume transfer
between CPU and GPU, a cost that increases further when in parallel, as the current
implementation of IndeX triggers re-execution of the data I/O due to larger than
usual ghost layer requirements. Work is in progress\footnote{personal communication
with NVIDIA Dev. team} to minimize this impact in a future version of the plugin.
\subsection{CPU-based rendering on a single node}
If memory costs are substantial, more nodes, and/or more GPUs will be required,
increasing the run-time cost of the visualization. Our data domain is quite large, and we are not able to load a half-size domain on a single GPU node. Indeed, both
the 64 GB RAM on the node and the 16 GB RAM on the GPU are hard limitations.
The OSPRay-based CPU rendering is one way to alleviate this problem. We can load the full size
domain on a single node of the multi-core partition of Piz Daint with dual-Xeon
chips and 128 GB of RAM. We measured again the startup cost for the first image
at full HD resolution (1920x1080 pixels),
using 72 execution threads and found them to increase linearly with grid dimensions.
We tested the quarter-size, half-size and the full domain and report the delivery
of the first image in 1.07, 1.50, and 2.33 s, respectively. The associated cost
in RAM is also linear, at 18.4 GB, 36.5 GB and 73 GB, respectively. Of great interest
is OSPRay's management of memory. OSPRay volumes can be stored in two different manners.
The first variant named \it{shared structured volume} \rm matches ParaView's data layout.
Version 5.6 of ParaView is the first version where this zero-copy access pattern is used
and it provides both a faster startup time and a much lower memory footprint, as compared
to previous work. Indeed, we reported earlier on
the use of OSPRay's alternate implementation called \it{block bricked volume} \rm whereby
data locality in memory is increased by arranging voxel data in smaller chunks. This came
however at a higher cost, doubling the memory footprint on the CPU \cite{fav18b}. 
 
After the first frame has been built, our experience is that smooth interaction
is possible with all three libraries tested. In fact, ParaView supports acceleration
shortcuts for lower precision renderings during interactive navigation,
enabling a comfortable user experience for mouse-driven interaction,
with little degradation of quality. 
Color and opacity transfer functions editing is also interactive and very intuitive.

Movie quality renderings on the other hand are done with all level-of-details
optimizations turned off and we tested the rendering speed of that particular
mode in a batch production test.
We created an OSPRay-based benchmark test to mimic a navigation fly-through in
a full resolution domain, starting from an overall view of the full grid, zooming in, rotating
the view-point, and finally zooming in to immerse the viewer in the volume. Our initial view-point has some regions of screen-space empty, where rendering costs at each pixel are negligible. We then move quickly into the scene such that
the viewport is completely covered by active pixels, that is, all pixel rays hit the volume. We rendered our benchmark test at three different pixel resolution, WXGA (1280x800 pixels), Full HD (1920x1080 pixels)
and 4K Ultra HD (3840x2160 pixels), to evaluate the impact of pixel resolution on rendering costs.
We also evaluated the use of hyper-threading to further boost performance.
Table \ref{tab:osprayThreads} summarizes
our average rendering time per frame for 300 frames of navigation.

\begin{table}[htb]
  \centering
  \caption{
    Average rendering for the full size domain at different pixel resolution
  }
  \label{tab:osprayThreads}

  \begin{tabular}{lccc}
    \hline
    Pixel Resolution vs. \# of threads  & 36 threads & 72 threads\\
    \hline
    WXGA (1280x800 pixels) & 2.85 s &  1.69 s \\
    FHD(1920x1080 pixels) & 3.37 s &  1.90 s \\
    4K UHD (3840x2160 pixels) & 4.81 s &  2.73 s \\
    \hline

  \end{tabular}
\end{table}

Our compute nodes are featuring two sockets with eighteen cores each. We note
the clear benefit of using hyper-threading to spawn up to seventy-two threads
for an increased throughput. We also note that increasing frame buffer resolution
to very large sizes is not a showstopper. 

\subsection{Rendering the full domain in parallel}

\begin{figure}
	\centering
	\includegraphics[width=\linewidth]{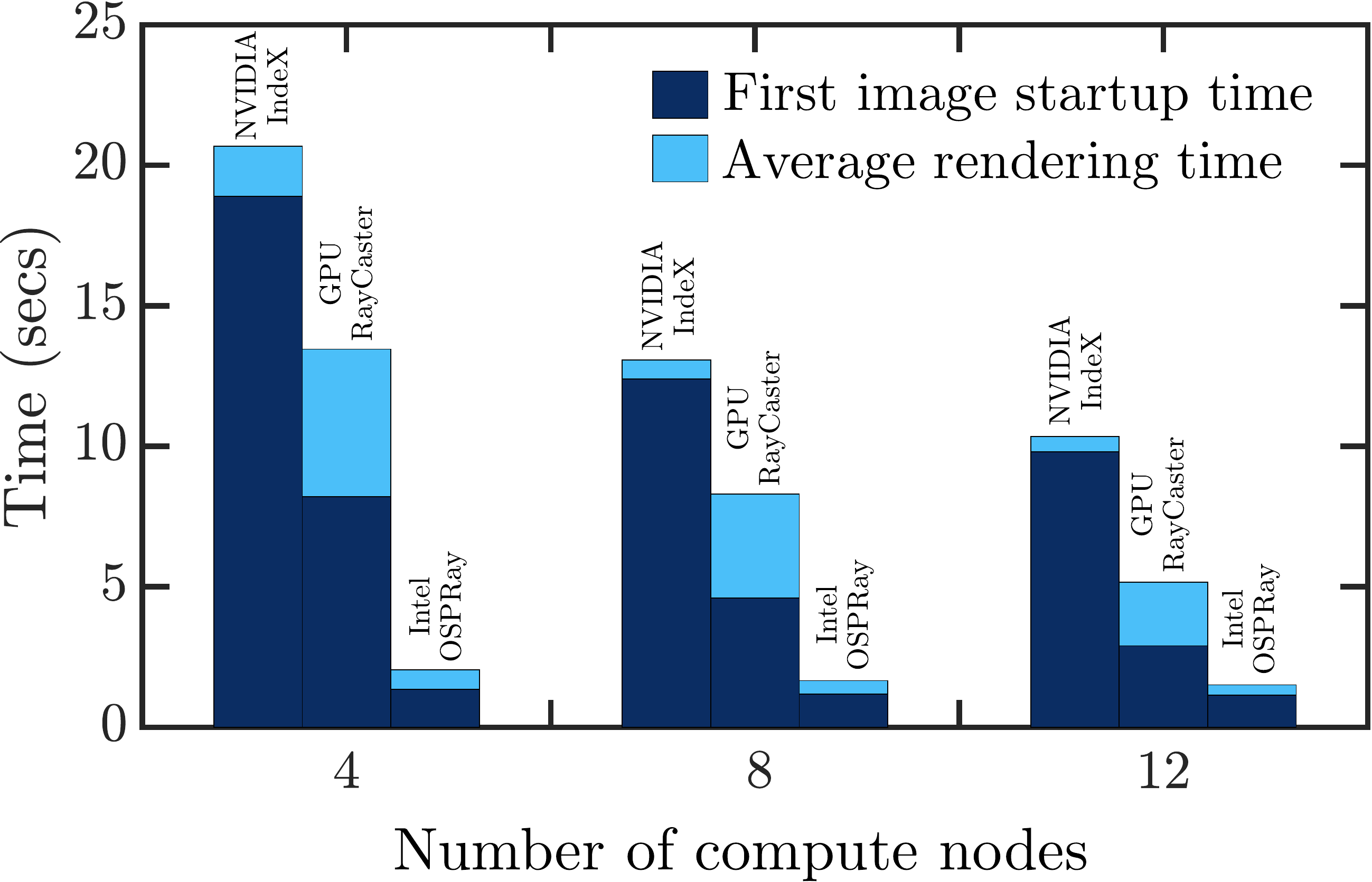}%
	\caption{\label{fig:diagram} Overview of initial cost and average rendering time per frame.}
\end{figure}

In a post-processing scenario, we have seen that the two GPU-based
rendering solutions are limited by the available GPU memory, since our 9-billion
voxels data set will not fit on a single GPU. Likewise, in an \it{in-situ} \rm
scenario, the visualization would most likely use a parallel 
set of nodes. Loading our full-size data, we rounded up our evaluation of all
three rendering options, by measuring the initial cost
for the first image (after all I/O has been done), and also the average rendering time
in a scripted animation loop. Fig.~\ref{fig:diagram} summarizes our results,
with the dataset distributed among 4, 8 and 12 compute nodes.


As expected, startup times decrease almost linearly with the number of compute nodes.
For the GPU-based methods, less data is transferred from CPU memory to GPU memory.
Our animation benchmark loads a single timestep of data, thus, once the data has migrated to the GPU, there is
hardly any CPU to GPU communication apart from a single frame buffer image.
For the CPU-based implementation, the build-up of the ray-tracing
acceleration structures takes just over one second so there is less difference across the few tests executed.
We see rendering times reduced somewhat linearly since there is less workload. 
In a movie production setting where all timestep outputs are read once, rendered once
and then discarded, the startup cost of any rendering library needs to be weighted
against the I/O costs. Although our data I/O statistics show quite a bit of variation
because of the high load of our multi user system with over 5000 compute nodes,
our simulation data are read, \it{in average}\rm, in about 32 s (resp.~25, and 16 s)
on 4 nodes (resp.~8 and 12 nodes). We see that the initialization of the rendering sub-system
has a greater impact than expected, and that in an \it{in-situ} \rm scenario,
it would be the singlemost important barrier to performance. The initialization of
the NVIDIA IndeX is the most significant bottleneck. Discussions with NVIDIA
are on-going and our hope is that this will be improved in future versions of the
SDK since the library is still in early development. We comment
here that the parallel execution of the OSPRay-based volume rendering was made possible
by using yet another ParaView mode, letting the OSPRay library take full control
of the overall scene and parallel frame compositing. Finally, we highlight the fact
that the OSPRay average rendering times per frame in our animation are all under one second,
while it takes a minimum of 8 compute nodes using the NVIDIA IndeX solution. This level of interactivity
can be satisfactory during the prototyping phase of a visualization.

\section{Summary and conclusion}

We have discussed three implementations of volume rendering for a thermal convection
simulation output of substantial size. Our time-dependent output is stored as a
\it{float32} \rm array of 36GB per timestep. This is a non-trivial size for the most common
GPUs. This leaves the scientist with two options: 1) use a data-parallel visualization
application with GPU-assisted rendering, or 2) use a \it{CPU-only} \rm
visualization environment which can fit on compute nodes where large memory
banks are usually found. Our choice was to deploy a single application, the open-source
ParaView, due to its support for different parallel execution paradigms, and for its ability to work with different off-screen and on-screen rendering backends. Having a single application,
driven by fully automatized python scripts and a benchmarking suite of tools
available in ParaView itself, enabled us to confront all possible implementations with reduced variability.

We tested two GPU-based rendering options. We first used ParaView's native volume rendering which has proved to offer the best compromise between startup time, and interactive performance; We also tested an alternative solution based on a new library in development by NVIDIA. In our current setup, the IndeX library offers superior interactive rendering, however at non-negligible initialization costs.

We evaluated an implementation of volume rendering provided by the Intel OSPRay library,
a software-based framework which can take remarkable advantage of a multi-threaded
execution layer. This also fits well on a subset of our available hardware, a dual-Xeon based compute node without GPU. Our experiences are of interest for several computer platforms around the world where graphics hardware is not available. 

Our emphasis on creating the scientific visualization shown in the accompanying video \cite{fav18a} was two-fold. 
First, having an interactive environment enabling us to prototype the visualization with large scale data. The editing of color and opacity transfer functions is the most demanding step in
deriving the proper visualization, and we were able to provide an interactive setup using either GPU-, or CPU-based volume rendering. Dealing with long time-dependent simulation outputs was the second requirement, and the path to achieve high productivity was to use parallel and scalable I/O routines. We used VTK's native XML partitioned file format convention for cartesian image data. This was pivotal for a quick turn-around time. The OSPRay-based implementation had the best performance in both initialization and average rendering time, but suffered from some parallel image compositing artifacts at inter-process boundaries. Given the very high spatial resolution of our grid, these artifacts are only visible at extreme zooming in the vicinity of ghost-cells between MPI-distributed data. To conclude and ensure the best visual quality, the compromise for movie production was to use small subsets of GPU nodes with ParaView's native volume renderer.

The volume rendering benchmarking platform deployed to analyze our large grid simulations provides a unique chance to observe sheared thermal convection in a very simple system with far reaching consequences. Furthermore, the visualizations allow us to have a very good first insight into the interplay between thermal convection and flow shearing by different kinds of wind and flow currents. We are now able to better understand the emergence and behavior of flow structures transporting heat through the system and affecting the flow dynamics.



\section*{Acknowledgments}

Alexander Blass was financially supported by the Dutch Organization for Scientific Research (NWO-I) and conducted his simulations at the Swiss National Supercomputing Center, under compute allocations s713, s802, and s874. We acknowledge the support from the Dutch national e-infrastructure of SURFsara, a subsidiary of the SURF cooperation, and the Priority Programme SPP 1881 Turbulent Superstructures of the Deutsche Forschungsgemeinschaft. We thank the ParaView development team at Kitware, USA, for fruitful discussions and motivational material. Dave DeMarle has been particularly helpful in discussion related to the OSPRay plugin. Mahendra Roopa at NVIDIA has also been extremely receptive to our feedback and instrumental in helping us get the best of the IndeX library in a multi-GPU setting. We are grateful to the reviewers of our manuscript who provided critical reading and motivated clarifications we have added. We also would like to thank Paul Melis from SURFsara for valuable input to our video \cite{fav18a}. 

\bibliographystyle{elsarticle-num} 
\bibliography{bibliography-template}

\end{document}